\documentclass[sigplan,10pt,nonacm]{acmart}
\usepackage{ifthen}
\usepackage{fancyhdr}
\usepackage{xspace}
\usepackage{booktabs}
\usepackage{mathptmx}   
\usepackage{courier}
\usepackage{color}
\usepackage{graphicx}
\usepackage{multicol}
\usepackage{multirow}
\usepackage{adjustbox}
\usepackage{verbatim}
\usepackage{appendix}
\usepackage{titlesec}

\newcommand{\TITLE}{Enabling Security-Oriented Orchestration of Microservices}

\newcommand{\PAGENUMBERS}{yes}

\newcommand{\ANONYMOUS}{no}
\newcommand{\SHOWTOAPPEAR}{no}
\newcommand{\COMMENTS}{no}

\usepackage[absolute]{textpos}
\newcommand{\ToAppear}{%
\begin{textblock*}{\textwidth}(0.95in,0.4in)
\begin{flushright}
    \noindent{\fbox{\textsf{Draft version---please do not redistribute.}}}
\end{flushright}
\end{textblock*}
}

\usepackage{caption}
\usepackage{algorithm}
\usepackage[noend]{algpseudocode}

\usepackage{listings}

\newcommand{\eg}{e.g.,\xspace}

\newcommand{\ea}{\emph{et al.}\xspace}
\newcommand{\Parabreak}{1.5ex}
\newcommand{\Paragraph}[1]{\vspace{\Parabreak}\noindent\textbf{#1}}
\newcommand{\swsupchain}{software supply chain\xspace}

\newcommand{\sgx}{Intel SGX\xspace}

\newcommand{\ignore}[1]{}

\ifthenelse{\equal{\COMMENTS}{yes}}{%
\newcommand{\msm}[1]{\textbf{MSM: #1}}
} {
\newcommand{\msm}[1]{}
}

\usepackage{enumitem}
\setlist[itemize]{noitemsep,nolistsep}
\setlist[enumerate]{noitemsep,nolistsep}

\usepackage{colortbl}
\definecolor{Gray}{gray}{0.9}
\definecolor{LightGreen}{rgb}{0.2,1,0.4}

\usepackage{longtable}

\date{}
\title{\TITLE}
\ifthenelse{\equal{\ANONYMOUS}{no}}
{
\author{Marcela S. Melara}
\affiliation{%
  \institution{Intel Labs}
  \city{Hillsboro}
  \state{OR}
  \country{}}
\email{marcela.melara@intel.com}

\author{Mic Bowman}
\affiliation{%
  \institution{Intel Labs}
  \city{Hillsboro}
  \state{OR}
  \country{}}
\email{mic.bowman@intel.com}
}{}

\ifthenelse{\equal{\PAGENUMBERS}{yes}}{%
  \pagestyle{plain}
}{%
  \pagestyle{empty}
}

\begin{document}
\begin{abstract}
As cloud providers push multi-tenancy to new levels to 
meet growing scalability demands, ensuring that externally developed
untrusted microservices will preserve tenant isolation has become a high priority.
Developers, in turn, lack a means for expressing and automatically 
enforcing high-level application security
requirements at deployment time. %

In this paper, we observe that orchestration systems are ideally situated 
between developers and the cloud provider to address these issues.
We propose a security policy framework
that enables \emph{security-oriented} orchestration
of microservices by capturing and auditing code properties that are
incorporated into microservice code throughout the \swsupchain. 
Orchestrators can leverage these properties to deploy microservices on a node 
that matches both the developer's and cloud provider's security policy \emph{and} their 
resource requirements.
We demonstrate our approach with a proof-of-concept
based on the Private Data Objects~\cite{pdo-git} confidential smart contract framework,
deploying code only after checking its provenance. 
\end{abstract}

\maketitle

\ifthenelse{\equal{\SHOWTOAPPEAR}{yes}}{\ToAppear}{}

\section{Introduction}
\label{sec:intro}
Microservices are emerging as a dominant software design paradigm for 
distributed applications easing deployment effort for developers, 
and enabling cloud providers to meet increasing resource demands while
providing component-level fault isolation.
However, three major trends in distributed computing introduce new security
challenges for deployment.

\textbf{1. Shift towards lightweight containers.}
Cloud providers improve microservice performance and increase multi-tenancy
using lighter weight, finer grained execution environments that rely on software-based techniques for
security and isolation (\eg~\cite{faasm,fastly-lucet}).
However, this means cloud providers can no longer rely on VM-like
mechanisms to protect their resources and strongly isolate co-tenant microservices.

\textbf{2. Multiplicity of principals and locations.}
Distributed applications are becoming increasingly data-centric
keeping data stationary as costs to move large troves of data 
have become prohibitive. 
This shift often results in data being spread across multiple 
organizations (\eg for federated learning or decentralized data analytics) 
and locations (\eg cloud and edge). 
Thus, microservice deployments may span multiple geographically 
and institutionally distributed compute resources, each with their 
own security features and requirements.

\textbf{3. On-demand deployment.}
Cloud providers are primarily relying on orchestration frameworks to select on-demand 
which compute nodes are most suitable to deploy particular microservices. 
Yet, today's orchestrators (\eg~\cite{kubernetes,aws-ecs,openstack-heat,docker-swarm}) 
select nodes optimizing for compute resource usage, but do not consider 
microservice security in this decision.

\Paragraph{The core problem.}
Microservice deployments must adhere to a multitude of security requirements 
imposed by mutually distrusting \emph{application principals}-- software developers,
data owners, and cloud providers. 
Although application principals wish to enforce their individual security requirements,
they do not currently have a common way of easily identifying, expressing and automatically 
enforcing these requirements end-to-end from source code to deployment.

Because cloud providers have intimate knowledge of 
available compute platforms they can specify \emph{fine-grained} requirements about unknown 
microservice code they deploy (\eg the code must provide memory bounds checks).
In contrast, developers/data owners may not have a priori knowledge 
about the platform hosting their code expressing only \emph{coarse-grained}
policies (\eg the code must be deployed in a sandboxed execution environment).

Since orchestration systems interpose on microservice deployments
for scheduling and resource management, we see an opportunity to leverage 
their central position in front of the compute nodes at the cloud provider to 
address this issue. 

\Paragraph{Our Proposal.}
We introduce the concept of \emph{security-oriented orchestration}: 
Selecting a suitable host platform for a workload requires evaluating multi-principal security policies, 
and ensuring that the execution environment preserves all principals' security requirements in addition
to meeting performance and latency requirements.
We propose CDI (Code Deployment Integrity), a security policy framework designed to
enable such security-oriented orchestration.

Our key insight is that even if application principals do not trust each other directly,
they can establish trust in a microservice 
\emph{if they trust who was involved in creating the final executable} that is deployed at 
a cloud provider. This is possible since microservice-based deployments rely very heavily on a complex 
\swsupchain to preserve certain code security properties during the creation of a microservice.
That is, even if the code is transformed or inspected through various tools, if those
tools operate as expected, a final executable microservice is expected to provide
specific properties (\eg compilation inserts stack canaries 
for memory safety).

Thus, CDI captures properties about the operations performed throughout the \swsupchain
on a piece of software and binds this metadata to the artifact in a cryptographically 
verifiable manner. 
CDI then uses this metadata to establish an auditable \emph{provenance chain} between 
the developer's source code and the executable microservice that the orchestrator receives for 
deployment.

By evaluating the CDI metadata attached to a microservice, an orchestrator 
can perform two security-oriented tasks. First, it can map 
cloud provider requirements to developer or data owner policies bridging the gap
between coarse and fine-grained security requirements. Second, the orchestrator can use the metadata to identify
an execution environment that minimizes performance costs (\eg a lightweight runtime) 
while trusting the microservice to meet all security requirements (\eg the code is sandboxed).

However, a plethora of attacks have undermined the integrity 
of entire applications by compromising one or more operations of the 
\swsupchain~\cite{solarwinds-fireeye,sgx-signing-injection,backstabbers-knife,left-pad-blog,left-pad-ars,poison-sw-supply,maloss,vulns-ci, sec-ci,cloud-assurance,off-my-cloud}.
To protect this vital ecosystem, our design provides integrity for the supply chain by
employing trusted execution environments (TEEs)
(\eg~\cite{sgx,sgx-paper,amd-sev,tdx}) for their hardware-enforced code 
integrity and authentication features.

Crucially, CDI does \emph{not} directly evaluate or guarantee the correctness or security
of microservice code. Instead, the provenance information obtained
throughout the \swsupchain provides an orchestrator with the 
\emph{history of transformations and inspections} performed to create a given 
microservice. Because of the code properties associated
with specific supply chain operations, as long as the orchestrator can establish trust in 
the provenance of a microservice, it can gain assurances about the security properties of the code by
proxy of its history. 
 
We demonstrate our approach with a proof-of-concept 
based on the \sgx~\cite{sgx,sgx-paper} TEE and the Private Data 
Objects~\cite{pdo-git} confidential smart contract framework,
which deploys code only after verifying its provenance.
\section{Related Work}
\label{sec:related}
\Paragraph{Orchestration and security.}
Today's orchestration frameworks (\eg~\cite{k8s-sec,aws-ecs-sec,docker-sec,openstack-sec})
largely place the burden of enforcing application security requirements on developers 
through complex, low-level per-container configuration.
Kubernetes~\cite{kubectl-plugins} allows cloud providers to advertise available \emph{platform-specific} security 
features such as TEEs, but developers must still reason about each hardware feature and manually request
a specific platform in their deployment configuration.

Several frameworks~\cite{kritis,binary-auth,docker-sec,openstack-sec} 
provide the ability to only deploy verified trusted container images.
These approaches share techniques with CDI, providing access control
to the platform and vulnerability-based security policy enforcement, and could be used in
combination with CDI to build a more security-oriented orchestration system.

\Paragraph{Supply chain security.}
Cloud providers~\cite{grafeas,container-analysis} and academics~\cite{in-toto} 
have proposed solutions that offer end-to-end security metadata collection and
policy enforcement throughout the \swsupchain. 
However, these approaches do not easily enable an orchestrator to leverage
these policies to make security decisions about the deployment itself.

Concurrent work in Craciun \ea~\cite{sgx-signing-injection} proposes mitigations for attacks 
that circumvent \sgx integrity checking by injecting malicious code into the 
enclave binary between the enclave build and signing stages.
Since the proposed approach shares techniques with CDI, we imagine these mitigations may be
used in combination with CDI in confidential computing scenarios.
\section{Overview}
\label{sec:design-goals}

Our goal for CDI (Code Deployment Integrity) is
to enable security-oriented microservice orchestration
that enforces security policies imposed by different 
application principals: cloud providers, software developers, and data owners.

\subsection{Terminology}
The \swsupchain is a directed acyclic graph (DAG) of \emph{operations} performed to create 
a deployable \emph{executable bundle}.
An operation by a supply chain \emph{tool} transforms or inspects one or more input
\emph{software artifacts} generating one or more output artifacts, which may in turn be
passed in as input artifacts to subsequent \swsupchain operations. Thus, artifacts
comprise materials such as source code files, intermediate code representation such as bytecode, 
configuration files, test results, log files, executable binaries, shared libraries, 
a Docker image, or Python package.

As CDI collects \emph{security metadata} about the transformation or inspection performed
by a given tool for later evaluation by an orchestrator, this metadata is attached to the
generated artifacts essentially providing an annotated \swsupchain DAG, which serves as an 
auditable trace.
Thus, we refer to an executable bundle as an artifact that can be deployed at a cloud provider
with the attached CDI metadata.

\subsection{Principals \& Assumptions}
To optimize resource usage, \emph{cloud providers} may run application code in a 
multi-tenant environment, requiring hosts to protect applications from one another,
while also appropriately protecting their own resources against unauthorized access
or tampering.
At the same time, a \emph{developer} may want to ensure that their microservice 
is deployed in a way that meets their application security policies and configuration.
However, we assume that developers may not know a priori the specific compute 
platform (and specifically its exact hardware features) their microservice will be deployed on.

In addition, we treat the data owner as a separate application principal;
owners of large-scale datasets may for example loan out
access to the data (or a subset thereof) to software developers
wishing to run analytics on the data. 
We also assume that the data owner may not fully trust the developer,
imposing confidentiality, usage or regulatory constraints on the data that are 
expected to be enforced at deployment time. 

Since microservices are created through a complex \swsupchain, application
principals expect each tool that transforms or inspects an artifact to
preserving specific security properties in the code as part of its operation 
However, because the cloud provider or software developer may each be responsible 
for configuring (parts of) the \swsupchain for a given microservice, we assume
individual tools may not be fully trusted by one of the participants.

\subsection{Design Goals}
The design of CDI meets the following three goals.

\Paragraph{G1: Any application principal must be able to validate
the full provenance of the code.} By digitally signing the CDI
metadata bundled with a microservice at each stage of the \swsupchain, 
CDI provides an auditable trace of the operations
a piece of code has undergone from its source at the developer to its executable
form at the cloud provider, attributing each stage to the responsible
tool.

\Paragraph{G2: Orchestrators can establish trust in the code at deployment time
without a priori knowledge of the tools that created it.}
CDI achieves this goal by enabling independent \emph{vetting authorities} to validate 
and certify that specific tools properly provide the expected security properties.
Then, as long as orchestrators are able to establish a chain of trust between a
vetting authority and a given tool, the signed CDI metadata is sufficient
for the orchestrator to dynamically gain trust in the deployed microservice.

\Paragraph{G3: Application principals need not rely on a single centralized
root of trust.}  
Developers, data owners and cloud providers may independently select the vetting
authorities they trust to properly certify tools in the \swsupchain ecosystem.
A security-oriented orchestrator, then uses this information when enforcing the principals'
policies.
\section{CDI Design}
\label{sec:cdi-design}

At a high level, CDI enables principals in 
distributed applications to express security policies, and
allows an orchestrator to enforce these policies through 
metadata about the \swsupchain. 
In CDI, \emph{vetting authorities} are entities that independently vet and 
digitally certify that a given tool performs a transformation
(\eg compilation) or inspection (\eg unit testing) that preserves
specific code security properties in the microservice bundle (see~\S\ref{secsec:vas}).

To capture these properties at every \swsupchain operation, 
each tool generates a digitally signed
statement summarizing its operation over the artifacts, which we call
a \emph{CDI report} (see~\S\ref{secsec:cdi-reports}).
Dedicated \emph{policy engines}, which can be incorporated into
an orchestration framework, then evaluate the CDI reports to verify
the provenance of the microservice and ensure the deployment will adhere
to all principals’ security policies \emph{prior} to execution.

For example, if the cloud provider has a WebAssembly (Wasm)-based lightweight
container available (\eg as in~\cite{fastly}) and wants to ensure that deployed 
microservices cannot violate software-based isolation, the CDI report allows the 
orchestrator to check if a microservice was created by a Wasm build toolchain that 
the host trusts to preserve the Wasm-mandated memory bounds checks needed for code 
sandboxing. Then, if the orchestrator can establish trust in the provenance of the 
code, it can confidently deploy the microservice in the more efficient execution environment.
Meanwhile, the microservice developer only needed to specify a policy requesting
deployment in a Wasm-based container.

\begin{figure*}[t]
  \centering
    \includegraphics[width=0.8\textwidth]{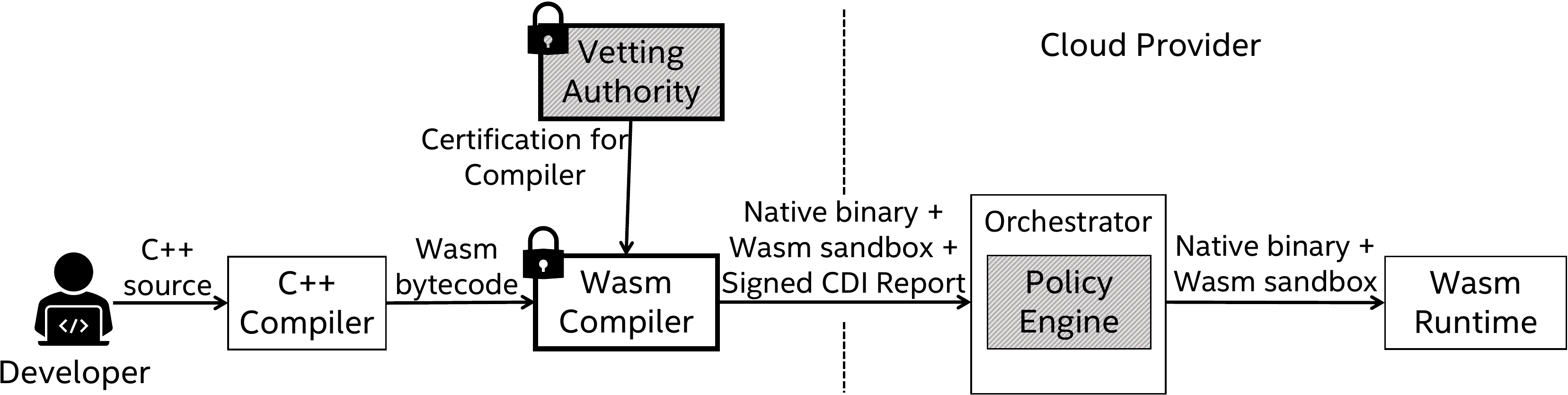}
    \caption{\label{fig:cdi-arch} \textbf{CDI architecture for an end-to-end \swsupchain for a Wasm-based deployment. The shaded gray boxes depict the two new components CDI introduces into the pipeline, Vetting Authorities and Policy Engines. Components depicted with the lock symbol are run inside of a TEE.}}
\end{figure*}

To protect the integrity of vetting authorities and individual tools,
CDI employs trusted execution environment (TEE) technology such as
\sgx~\cite{sgx,sgx-paper} for its hardware-enforced security properties (see~\S\ref{secsec:tee}).
We summarize the CDI architecture for a simple Wasm-based supply chain in Fig.~\ref{fig:cdi-arch}.

\subsection{Vetting Authorities}
\label{secsec:vas}
To avoid requiring application principals to have full knowledge of the
\swsupchain prior to deployment, CDI vetting authorities 
act as trusted intermediaries between application
principals and individual supply chain tools.
Specifically, vetting authorities are responsible for certifying that
a given tool properly implements or preserves specific security properties in
the artifact, \eg \texttt{gcc} inserts stack canaries into compiled binaries for memory safety.

CDI requires vetting authorities to produce a tool certification $C$ signed with 
the vetting authority's signing key $SK_{VA}$ indicating the security properties 
provided by a given tool:
$$\text{C}=\text{Sign}_{SK_{VA}}\left(T || P \right)$$
where $T$ is a representation of the tool's attributes (\eg name, version, tool owner's release signing key)
including its public report verifying key $PK_T$ and $P$ a string-representation of the certified
properties later used for policy enforcement.

In practice, we expect each vetting authority to determine the exact process
by which to vet individual \swsupchain tools. 
This vetting process may be as basic as asserting institutional trust in
a particular tool or as complex as performing formal verification of tool.
However, given the large number of \swsupchain tools available to developers and
hosting services today, our design assumes that a number different of vetting
authorities will be needed to vet different types of tools. 

Thus, our design supports creating a hierarchy of vetting authorities
akin to how traditional certificate authorities operate today.
A CDI vetting authority $VA$ may thus also certify other vetting authorities
establishing a chain of trust between \emph{root} vetting authorities and
tool vetting authorities via a signature chain $SIG_{VA}$,
consisting of the following fields:
$$\text{SIG}_{VA}=\text{Sign}_{SK_{VA}}\left(PK_{c} || \text{SIG}_{p} \right)$$
where $PK_c$ is the child vetting authority's public verifying key used to identify
the child vetting authority and $SIG_p$ is the parent vetting authority's signature
chain including the signature on $VA$'s public key $PK_{VA}$. If applicable, 
a tool certification $C$ will include the vetting authority's $SIG_{VA}$.

\subsection{CDI Reports}
\label{secsec:cdi-reports}
CDI collects verifiable provenance information about a microservice by
generating a CDI report at each \swsupchain tool.
Each report captures metadata about operation the specific tool performed
on its input artifacts to create the output artifacts.

Specifically, after each tool $t$ performs its operation, the tool
signs its output, as well as operation-specific metadata, using their report signing key $SK_t$. 
Specifically, a CDI report $R$ consists of:
$$\text{R}=\text{Sign}_{SK_t}\left(C || \text{SIG}_{VA} || M || \text{H}(\text{output}) || \text{H}(\text{R}_{in}) \right)$$
where $C$ is a vetting authority certification of the tool and
$SIG_{VA}$ is tool's vetting authority signature chain. $M$ is additional metadata on the
operation, which may include configuration information (\eg compiler flags used),
cryptographic algorithms in use, etc.
Since the tool output may be very large (\eg entire compiled binary),
a CDI report includes the cryptographic hash of the output, which
also provides an additional layer of integrity for the output.


If available, $R$ also includes the hash of each CDI report $R_{in}$ attached to
each input artifact generated by ``upstream'' tools creating a hash tree 
that forms a linked \emph{provenance chain} for the entire \swsupchain DAG.
This hash tree allows policy engines to
validate the end-to-end integrity of the \swsupchain, as well
as obtain a commitment that the microservice bundle produced at the end
of the \swsupchain corresponds to the artifact that entered the first operation
of the supply chain (\eg C++ source code).
In practice, validating the full supply chain hash tree may become too
costly for cloud providers or orchestrators; we envision allowing application
principals to customize provenance validation on a per-operation basis at the 
orchestrator, or providing a blockchain-based solution to
facilitate the retrieval of the provenance chain.



\subsection{Supply Chain Integrity}
\label{secsec:tee}
The emergence of TEE technologies offer an opportunity 
to strengthen the integrity of the \swsupchain and vetting authorities.
Though our approach in CDI is general, we describe a design
based on \sgx~\cite{sgx,sgx-paper}.
\sgx is designed to protect the confidentiality and
validate the integrity of application code and data, 
even in the face of an untrusted or compromised 
host by providing a \emph{hardware enclave}, an encrypted memory region
within the address space of a userspace process. 
To enable integrity checking of enclave code,
\sgx computes a digitally signed SHA-256 hash of the enclave
memory when the enclave is initialized. 
This signed hash also allows remote parties to authenticate the enclave code 
via the \sgx attestation protocol~\cite{sgx-epid}.

By running individual supply chain tools or components and vetting authorities 
in an \sgx enclave, CDI can gain three benefits.
First, we can preserve the confidentiality of privacy-sensitive data 
such as proprietary source code at build time or vetting authority cryptographic keys.
Second, vetting authorities can use \sgx remote attestation to 
ensure they are certifying the expected tool, while policy engines may remotely authenticate
trusted vetting authorities.
Third, \sgx can help CDI extend and enforce provenance in hardware as tools and vetting
authorities may derive their signing keys within an enclave, binding tool certifications
and CDI reports to the specific platform that generated the keys.
\section{Proof of Concept}
\label{sec:implementation}

We have implemented a prototype of CDI for the 
WebAssembly build toolchain for the Private Data Objects 
(PDO)~\cite{pdo-git} framework for confidential off-chain smart 
contracts.
To preserve data confidentiality, execution integrity and enforce data access
policies defined in the contract, PDO executes smart contracts inside \sgx-based
\emph{contract enclaves}.


\subsection{Background}
Smart contracts are programs that are used to automatically execute an agreement or protocol 
between multiple parties relying on a decentralized blockchain infrastructure
rather than trusted intermediaries for compliance with the agreement~\cite{smart-contracts-def}.
Though used for different purposes, smart contracts architecturally share many similarities 
with microservices: they typically implement a subset of functionality within a larger application,
they are deployed on-demand when an application participant requests a result, and
they may only be invoked through a small pre-defined API.

Private Data Objects (PDO) are designed for formalizing 
data access and update policies through smart contracts.
Developers can implement PDO contracts in any programming language
that can be compiled into WebAssembly. At run time, 
the PDO contract enclave dynamically loads and executes the deployed
contract code into the WebAssembly runtime running inside the enclave.

WebAssembly (or Wasm) is a binary format whose programming model
mandates built-in code sandboxing.
Since Wasm was initially designed 
for portable execution of code in browsers, being able to isolate different 
mutually distrusting modules running in the same runtime is a 
necessary security feature. Thus, the Wasm binary format requires that memory 
bounds checks to be inserted around modules to prevent cross-module access. 

The Wasm runtime used in PDO is based on the WebAssembly Micro Runtime (WAMR)~\cite{wamr},
which supports executing both Wasm bytecode as well as ahead-of-time compiled
binary code built by a WAMR-specific AoT compiler.
Due to the performance gains achieved by running AoT compiled Wasm binaries compared
to running interpreted Wasm bytecode, our prototype adds support for loading AoT compiled
smart contracts into a PDO contract enclave.

\begin{figure}[t]
  \centering
    \includegraphics[width=0.5\textwidth]{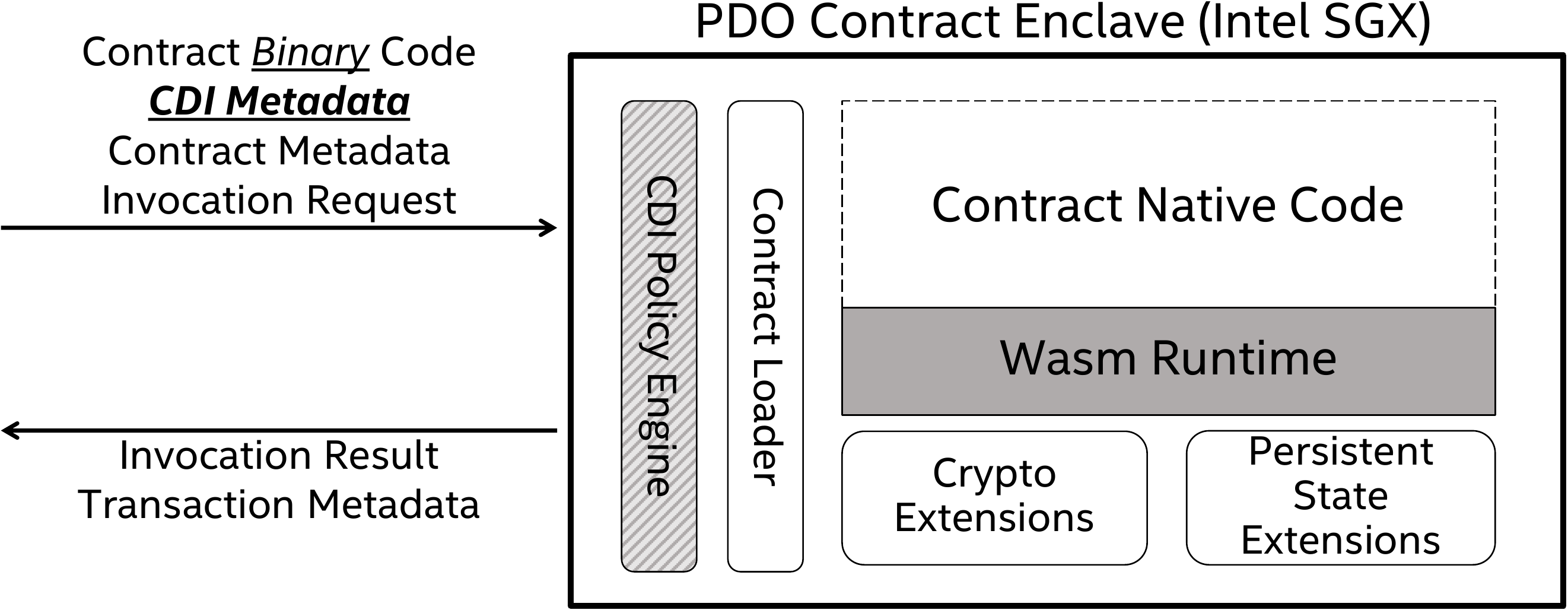}
    \caption{\label{fig:pdo-cdi-poc} \textbf{Overview of a PDO Contract Enclave for Wasm-based contracts with support for CDI. The Wasm Runtime (dark gray box) is provided by WAMR~\cite{wamr}, while the contract code (dashed box) is provided by the untrusted software developer.}}
\end{figure}

\subsection{Validating our Approach}
\label{secsec:eval}
PDO provides an ideal experimental platform for CDI for two reasons.
First, the PDO contract enclave represents an example of a lightweight isolated execution
environment used in emerging microservice-based deployments today. 
Second, Wasm's mandated memory bounds checks allow us to enforce the type of security 
properties that are expected by the cloud provider to be preserved by the \swsupchain, 
while enabling us to express coarse-grained developer-side security policies. 
Thus, we believe our CDI prototype demonstrates the feasibility of enabling deployment-time
trust establishment and security-oriented deployment through code provenance in a practical scenario.

\Paragraph{Experimental platform.}
To enable the WAMR AoT compiler to generate CDI reports about the compilation
of Wasm bytecode into Wasm binary, we built a wrapper around the compiler that
computes the SHA-256 hashes of the bytecode input and binary output, records any
compiler flags used, and captures these metadata in an ECDSA-signed CDI report.
For simplicity, we implemented a simple root vetting authority
as a PDO contract that generates an ECDSA signature on another ECDSA key
allowing us to take advantage of PDO's confidentiality and integrity
properties to protect the vetting authority's signing key.
Figure~\ref{fig:pdo-cdi-poc} describes the architecture of our prototype.
We discuss challenges and practical approaches to run our enhanced Wasm AoT compiler 
as well as other types of \swsupchain tools inside an \sgx enclave in~\S\ref{sec:future}.


In a preliminary performance evaluation,
we found that the additional security checks introduced by CDI do not
have a significant performance impact on vanilla PDO.
That is, the run time improvements we gain by running AoT-compiled 
smart contracts in PDO compared to running interpreted Wasm contracts
remain significant even with CDI enabled.

\Paragraph{Security Policy Specification.}
Our prototype assumes the cloud provider hosting PDO contract enclaves 
seek to enforce a single security policy: that AoT-compiled Wasm contracts 
originate from a Wasm build toolchain that preserves Wasm code sandboxing.
Thus, for simplicity our prototype accepts two types of host
security policies: a default \emph{accept-all}
policy, which essentially acts as a no-op, and a \emph{default-deny} policy which
requires that the provenance of all contracts must be validated before deployment. 
To avoid requiring a cloud provider to have a priori knowledge of every
possible Wasm build toolchain, a \emph{default-deny} policy must include
a set of trusted CDI vetting authorities.

Since PDO runs smart contracts inside of \sgx enclaves, we assume an
implicit developer policy that requires contracts to be run in a TEE.
To enable developers to verify that their contract is deployed
using \sgx, our prototype relies on the validation mechanism PDO
provides.


\Paragraph{Policy Enforcement.}
At contract enclave startup, an integrated CDI policy engine is initialized
with the host-specified security policy.
Our prototype enables a PDO contract enclave to make a security-oriented
deployment decision by validating the provenance of a smart contract bundle.

That is, the policy engine intercepts each incoming bundle,
and validates three properties using the attached CDI report: 1) That it can establish
a chain of trust between the vetting authority that certified the Wasm build toolchain
(in our case the WAMR AoT compiler) and a trusted root vetting authority. 
2) That the CDI report contained in the contract bundle was generated 
by this certified Wasm toolchain. 3) That the contract binary to be deployed matches
the binary indicated in the CDI report.
Only if these three properties hold can the contract enclave gain trust that the 
contract binary contains the required Wasm sandboxing properties
in order to safely deploy the contract.

\section{Discussion \& Future Work}
\label{sec:future}
Our proof-of-concept for CDI leaves many open challenges
that must be addressed to make our approach practical for
emerging distributed applications. 

\Paragraph{Creating a high-integrity supply chain with TEEs.}
Porting supply chain tools such as compilers to TEEs often introduces additional 
adoption complexities. For example, since \sgx provides a restricted interface to standard OS features,
running a tool such as the enhanced Wasm AoT compiler in our prototype would
require significant refactoring efforts.
One approach to address this issue is to use a TEE-aware LibraryOS (\eg~\cite{graphene, scone})
which supports running unmodified applications in the TEE.
Nevertheless, given the growing number of TEE technologies (\eg~\cite{amd-sev,tdx}), 
important future work includes evaluating a larger set of supply chain tools in TEEs, 
ensuring interoperability, and providing consistent provenance properties.

\Paragraph{Detecting inconsistencies.}
For application principals with stricter security requirements,
validating a single version of the full provenance chain may not be sufficient
to establish trust in a microservice bundle.
We envision addressing this issue with a protocol extension 
that allows principals to specify a threshold number of signatures by trusted vetting authorities
on specific supply chain operations to increase their confidence in the trustworthiness of a tool's
output. For example, a cloud provider may require that 3 out of 5 of its trusted vetting authorities
sign a Wasm compiler's report certifying that it preserves Wasm sandboxing.


\Paragraph{Making policy specification practical.}
CDI must ultimately enable data owners and developers to express 
\emph{concrete} yet platform-agnostic policies and to map these 
to fine-grained cloud provider requirements.
We have been exploring human-readable \emph{security tags}
such as \texttt{CODE\_SANDBOXING} or \\
\texttt{CONFIDENTIAL\_EXECUTION}, which can be mapped to security properties like
Wasm memory bounds checks in code or specific TEE hardware. 
However, since this approach would require identifying an exhaustive list of
enforceable security requirements, which is likely intractable, a more flexible and
expressive policy specification process is needed.

\Paragraph{Preserving security requirements in scheduling.}
Security-oriented orchestration ultimately requires 
the orchestrator to make scheduling decisions based on the performance
and security tradeoffs between different combinations of available hardware security features
(\eg VM vs. TEE) and target execution environments (\eg container vs. lightweight runtime).
In addition, the performance impact that certain higher-level security requirements may have on 
resource utilization (\eg single-tenancy for a workload) must be captured as part of the node selection decision. 
Identifying techniques to address this major challenge is a crucial next step.
\section{Summary}
We have introduced the concept of security-oriented orchestration,
which ensures that microservice deployments meet multi-principal security 
requirements. To enable such orchestration, we have presented CDI (Code Deployment Integrity)
a security policy framework that captures provenance metadata about the \swsupchain
that created a microservice.
By validating and establishing trust in the provenance of a microservice, CDI 
provides an orchestrator with assurances about the security properties of the code,
allowing it to enforce security policies prior to deployment.


\bibliographystyle{ACM-Reference-Format}
\bibliography{references}


\end{document}